\documentclass[%
 aip,
 amsmath,amssymb,
 reprint,
]{revtex4-1}

\usepackage{graphicx}
\usepackage[caption=false]{subfig}
\usepackage{dcolumn}
\usepackage{bm}

\usepackage[utf8]{inputenc}
\usepackage[T1]{fontenc}
\usepackage{mathptmx}

\usepackage{color}

\begin{document}

\title{Simulations of Non-Integer Upconversion in Resonant Six-Wave Scattering} 

\author{A. Griffith}
\email[]{arbg@princeton.edu}

\affiliation{Princeton University, Astrophysical Sciences}
\author{K. Qu}
\affiliation{Princeton University, Astrophysical Sciences}
\author{N. J. Fisch}
\affiliation{Princeton University, Astrophysical Sciences}

\date{\today}

\begin{abstract}
Resonant upconversion through a sixth order relativistic nonlinearity resulting in a unique resonance was recently proposed [V. M. Malkin and N. J. Fisch, Physical Review E 108, 045208 (2023)].
The high order resonance is a unique non-integer multiple of a driving pump frequency resulting in a frequency upshift by a factor of $\approx 3.73$.
We demonstrate the presence, unique requirements, and growth of this mode numerically.
Through tuning waves to high amplitude, in a mildly underdense plasma, the six-photon process may grow more than other non-resonant, but lower order processes.
The growth of the high frequency mode remains below the nonlinear growth regime.
However, extending current numerical results to more strongly coupled resonances with longer pulse propagation distances suggests a pathway to significant upconversion.
\end{abstract}


\maketitle 

\section{Introduction}
The falloff in available laser power below the optical is severe,~\cite{edwards_x-ray_2020-1} and producing high intensity UV, XUV, or X-ray pulses remains a challenge with current electron beam driven~\cite{emma_first_2010,pellegrini_physics_2016}  or laser driven sources.~\cite{powers_quasi-monoenergetic_2014,hort_high-flux_2019,edwards_x-ray_2020-1,chaulagain_eli_2022}
Recent proposals detail using resonant interactions in plasma to upconvert high fluence sources, such as the National Ignition Facility,~\cite{haynam_national_2007} to shorter wavelength.~\cite{malkin_towards_2020,malkin_resonant_2020-1,malkin_super-resonant_2022-1,griffith_modulation-slippage_2021-1}
This paper details a first numerical test of higher order plasma mediated wave mixing for upconversion.~\cite{malkin_six-photon_2023}

Wave mixing builds off of a long history of using plasmas components to replicate the behavior of well known solid state parts, but which operate at higher laser intensity. 
Pulse amplification can be done with electron plasma waves,~\cite{malkin_fast_1999,pai_backward_2008,ping_development_2009,trines_simulations_2011,edwards_beam_2017,qu_plasma_2017,miriam_cheriyan_comprehensive_2022} ion acoustic waves,~\cite{andreev_short_2006,weber_amplification_2013,edwards_x-ray_2017} or magnetized waves.~\cite{edwards_laser_2019}
Ion acoustic waves might also enable beam combination.~\cite{kirkwood_plasma_2018,kirkwood_plasma-based_2018,kirkwood_production_2022}
Plasmas could serve as polarizers,~\cite{michel_dynamic_2014,lehmann_plasma-based_2018}q-plates,~\cite{qu_plasma_2017} lenses,~\cite{edwards_holographic_2022} or gratings.~\cite{edwards_plasma_2022}

Mildly relativistic wave-mixing to upconvert optical light with high efficiency to much shorter wavelengths is a more recent proposal.
Pump photons are combined into a higher frequency output through nonlinear coupling.
The nonlinear coupling arises from corrections to the plasma frequency at significant electron Lorentz factor.~\cite{malkin_towards_2020}
This coupling contrasts with previous work in that it can be resonant, without requiring a modulated plasma density.~\cite{rax_third-harmonic_1992}
The lowest order nonlinearity results in four photon coupling.
A simple colinear alignment between pump and seed beams requires a complex set of resonance conditions~\cite{malkin_resonant_2020-1,malkin_super-resonant_2022-1} and non-colinear arrangements result in difficulties for amplification.~\cite{griffith_modulation-slippage_2021-1} 
A higher order, six-photon, process addresses both of these problems.~\cite{malkin_six-photon_2023}
This six photon scattering process  consumes four input photons with frequency $\omega$ to produce a pair of photons at frequencies of approximately $(2+\sqrt{3})\omega$ and $(2-\sqrt{3})\omega$.
This unique non-integer multiple frequency shift from the six photon process could allow for large changes in wavelength if it were to be cascaded multiple times.

We present a numerical investigation which demonstrates the unique six-photon resonance.
In contrast to previous work, here the resonance is not assumed in a reduced (slowly varying envelope) and expanded model.
Instead we consider a model, equivalent to a cold relativistic electron fluid, in which the full wave dynamics are evolved without the resonance embedded.
As the resonance is of higher order, coupling must be maximized to generate observable effects.
Thus, we consider the dynamics without expanding in either normalized plasma density, $n_0/n_c = \omega_{pe}^{-2}\omega^2$, for laser frequency $\omega$ and plasma frequency $\omega_{pe}^2 = 4\pi e^2 n_0m_e^{-1}$, or wave amplitude, $a = eA/mc^2$, for laser vector potential $A$.
When both parameters are not $\ll 1$, which is required to achieve significant coupling, there may be large errors in any expansion of nonlinear terms to finite order.
Numerical experimentation is used to examine the fundamental physics at play and validate the viability of relativistic six wave mixing for upconversion.

The exposition of our numerical results on relativistic six-wave mixing in plasmas is outlined as follows.
In Section \ref{sec:background} we review the dynamics and resonances available for the coupling between mildly relativistic waves, $a < 1$, in an underdense, $\omega_{pe} < \omega$ plasma.
In Section \ref{sec:numerics} we describe the choice of the particular numerical tools used to examine this regime.
In Section \ref{sec:results} we demonstrate the unique resonance proposed in by Malkin and Fisch.
In Section \ref{sec:conclusion} we discuss the applicability of this work to the aim of achieving significant upconversion, review our work, and discuss possible avenues for continued exploration.

\section{Background}
\label{sec:background}
We review and reformulate how the coupling of electromagnetic waves may be achieved through relativistic corrections to electron dynamics in strong electromagnetic waves.
The model in which the coupling occurs is that of an electron fluid which is driven purely by electromagnetic forces, neglecting electron temperature and pressure along with any ion dynamics.
Thus the fluid is strictly described by Maxwell's equations and the coupled electron quantities of density and velocity.
Working in the Coulomb gauge, the dynamics may be described by non-dimensionalized vector potential, $\mathbf{a} = e\mathbf{A}/m_ec^2$, electrostatic potential $\phi = e\Phi/m_ec^2$, and electron action $s = S/m_ec^2$ resulting in three coupled equations as written by Malkin and Fisch~\cite{malkin_resonant_2020-1}
\begin{align}
	\left[\partial_t^2 - c^2\partial_z^2 + \frac{\omega_{pe}^2 + c^2\partial_z^2\phi}{\sqrt{1+a^2+(c\partial_zs)^2}} \right] \mathbf{a} &= 0\label{eqn:a},\\
	\frac{\omega_{pe}^2 + c^2\partial_z^2\phi}{\sqrt{1+a^2+(c\partial_zs)^2}}  c\partial_z s + c\partial_t\partial_z \phi &= 0\label{eqn:phi},\\
	\partial_t s  -\phi + \sqrt{1+a^2+(c\partial_zs)^2} - 1&=0\label{eqn:s}.
\end{align}
In the case we shall consider, we assume propagation purely in the $z$ direction, with the vector potential $\mathbf{a}$ only having a transverse, $x$, component.
In this case the conservation of cannonical momentum makes the electron perpendicular momentum $p_\perp = mca$, and the action may be related to the more easily understood quantity of parallel momentum through the relation $p_z = mc^2 \partial_z s$.
To complete the correspondence between Equations \eqref{eqn:a}-\eqref{eqn:s} and cold electron hydrodynamics, electron density can be straightforwardly related to the electrostatic potential through Gauss's law, $\omega_{pe}^2 + c^2\partial_z^2\phi = \omega_{pe}^2 n_e/n_0$.

The dynamical equations contain multiple sources of scattering.
Raman scattering, where there is coupling between two electromagnetic waves and an electron plasma wave is the lowest order process.
When terms are kept to second order, fluctuations of density captured by the electrostatic potential in Equation \eqref{eqn:a}, which are driven by the ponderomotive force, derived from combining Equation \eqref{eqn:phi} and Equation \eqref{eqn:s} at the lowest order.
For the purposes of this paper, Raman scattering is purely parasitic, and forward Raman scattering may grow of off seeded fluctuations in the plasma to move energy into waves which are no longer resonant for upconversion.
The higher order processes responsible for coupling which may result in upconversion occur from the expansion of the electron Lorentz factor $\gamma = \sqrt{1+a^2+(c\partial_zs)^2}$.
The first term in this expansion gives the four-photon coupling~\cite{malkin_towards_2020} and the next provides the six-photon coupling.~\cite{malkin_six-photon_2023}
The subtleties between different resonance regimes and the precise scaling of the coupling in these regimes are not discussed here, as the particular expressions can be extensive and are best understood in a more thorough treatment.

The six-photon scattering process may be characterized by a simple set of resonance conditions and the asymptotic behavior of the coupling.
Perpendicularly polarized waves have dispersion relation, written as $\omega_j^2 = k_j^2 +n$ which is normalized for $k_0 = 1$ and $c=1$.
These waves may couple to one another through the relativistic nonlinearity.
The fifth order nonlinearity supports several different six-photon scattering processes, but the simplest case consists of one where four input photons at frequency $\omega_0$ and wavenubmer $k_0$ are turned into two photons, one at $\omega_1,k_1$ and one at $\omega_2,k_2$, meeting the resonance conditions
\begin{equation}
	4\omega_0 = \omega_1 + \omega_2, \qquad 4k_0 = k_1 + k_2.
\end{equation}
These conditions may be solved resulting in resonant frequencies
\begin{equation}
	\omega_1 = 2\omega_0 + \sqrt{3}, \qquad \omega_2 = 2\omega_0 - \sqrt{3}.
	\label{eqn:resonance}
\end{equation}
This resonance is unique in that it can occur for co-linear wavevectors in a plasma without renormalization of the laser frequency, which is not possible for lower order relativistic scattering.~\cite{malkin_resonant_2020-1}
Furthermore, in the limit, this produces a unique non-integer resonance, in contrast to typical integer harmonics.
The growth rate of the six photon coupling, $\gamma$,  normalized to the laser frequency, $\omega_0$, scales such that
\begin{equation}
	\gamma \propto (n/n_c)^3a_0^4\omega_0.
	\label{eqn:six_wave_growth}
\end{equation}
The form of the leading coefficient and the corrections in higher orders of plasma density may be seen in the exposition by Malkin and Fisch.~\cite{malkin_six-photon_2023}
From Eq. \eqref{eqn:six_wave_growth} it can be seen that the growth of any six-photon scattering is highly sensitive to both wave amplitude and density.

The parameters to produce a validation of the resonance outline by Eq. \eqref{eqn:resonance} are restrictive.
To validate the non-integer resonance, without assuming it a priori requires a full wave model.
This full wave model must have a high resolution to resolve the high frequency mode.
However, amplification distances resulting from the weak coupling given by Eq. \eqref{eqn:six_wave_growth} result in significant plasma and pulse lengths which are difficult to resolve.
To maximize the coupling, we use Eq. \eqref{eqn:six_wave_growth} to guide the parameter selection towards a high intensity $a_0$ and high plasma density $n$.
The more highly resolved model allows us to work in a regime where higher order corrections to the dispersion relation and to the equations of motion are more accurately captured than in a nonlinear series expansions or in an envelope based simulation.

\section{Numerical Methodology}
\label{sec:numerics}
\begin{figure}[t]
	\includegraphics[width=\columnwidth]{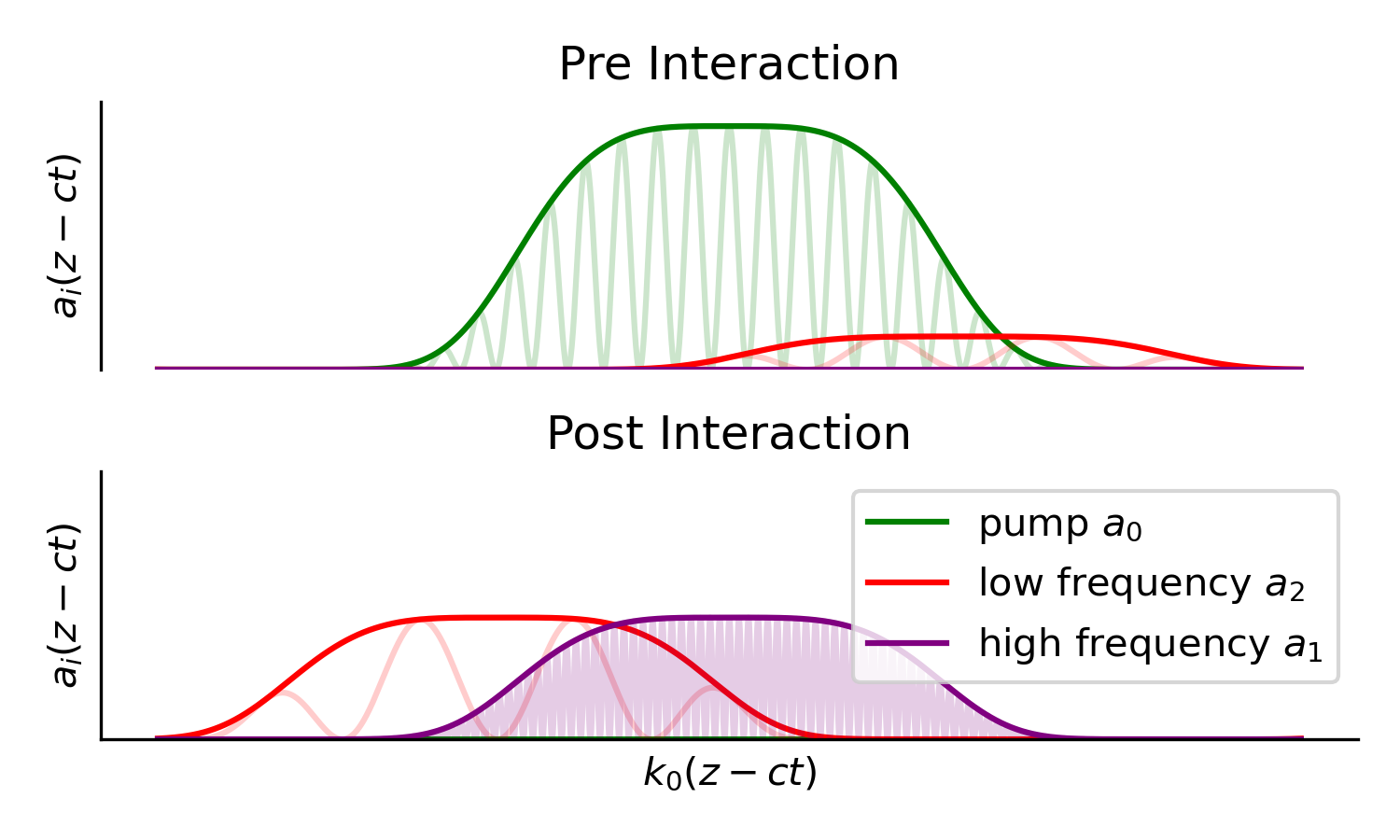}
	\caption{The pump wave is initialized with a super-Gaussian envelope which is led by a much weaker low frequency seed. If pump depletion could be reached, the ideal outcome would be such that the high frequency wave and low frequency wave are amplified until the pump is completely depleted. While at equivalent amplitude, the high frequency wave would contain the majority of the remaining energy due to having a much higher frequency.}
	\label{fig:diagram}
\end{figure}

The numerical evolution of the relativistic scattering process is chosen to maximize the observability of the six-photon scattering process.
To capture a distinguishable signal, the pseudospectral package Dedalus~\cite{burns_dedalus_2020} is used in order to minimize numerical noise in the Fourier spectrum, in contrast to high noise particle in cell results.
The nonlinear pulse propagation is non-dimensionalized and transformed into the co-propagating frame $\eta = t,\xi = z - t$:
\begin{align}
	\left[\partial_\eta^2 - 2\partial_\eta\partial_\xi + \frac{n_0 + \partial_\xi e}{\sqrt{1+a^2+l^2}} \right] a &= 0\label{eqn:num_a},\\
	 \partial_\eta e -\partial_\xi e+\frac{n_0+ \partial_\xi e}{\sqrt{1+a^2+l^2}} l&= 0\label{eqn:num_e},\\
	\partial_\eta l - \partial_\xi l  -e + \partial_\xi\left(\sqrt{1+a^2+l^2}\right)&=0\label{eqn:num_l},
\end{align}
where $c=1$ and density is normalized such that $n = \omega_{pe}^2/k_0^2$.
Given that only the derivatives of $\phi$ and $s$ couple to the other equation, for convenience we reduce the equations with the substitutions $\partial_z\phi = e$ and $c\partial_zs = l$.
To ensure numerical stability a viscosity term is added to curtail the rapid non-physical growth of high frequency modes driven by the complete absence of physical dissipation.

Simulations are initialized to test for the growth of the proposed six-photon scattering when waves are evolved by Eqs. \eqref{eqn:num_a}-\eqref{eqn:num_l}.
In a homogeneous plasma of density of $n$ we initialize a strong ``pump'' wave packet with $k_0 = 1$ and a  ``seed'' wave packet of varying $k_2$.
An example of the initial conditions can be seen in Fig. \ref{fig:diagram}.
The seed frequency is given a weak amplitude, and to account for different group velocities, is initially leading the pump wave.
In the ideal case, the pump wave is completely depleted, and all of the energy is in the low and high frequency waves post interaction.
The initial conditions for $a$, $e$, and $l$, are chosen such that the time derivative of each field are initially zero to the lowest non-zero order.
Through examination of the spectrum, and in particular the mode at $k_1 = 4k_0-k_2$ we demonstrate that Eqn. \eqref{eqn:resonance} describes a growing mode which is only seen when it is satisfied.

\section{Results}
\label{sec:results}
The presence of a uniquely resonantly growing mode can be demonstrated in the evolution of Eqns. \eqref{eqn:num_a} - Eqns. \eqref{eqn:num_l}.
For a strong laser pulse, many different scattering channels operate coincidently resulting in the growth of a large number of discrete modes.
The intensity of these modes is oscillatory, unless conditions are precisely chosen such to push waves into resonance.

\begin{figure}[t]
    \includegraphics[width=\columnwidth]{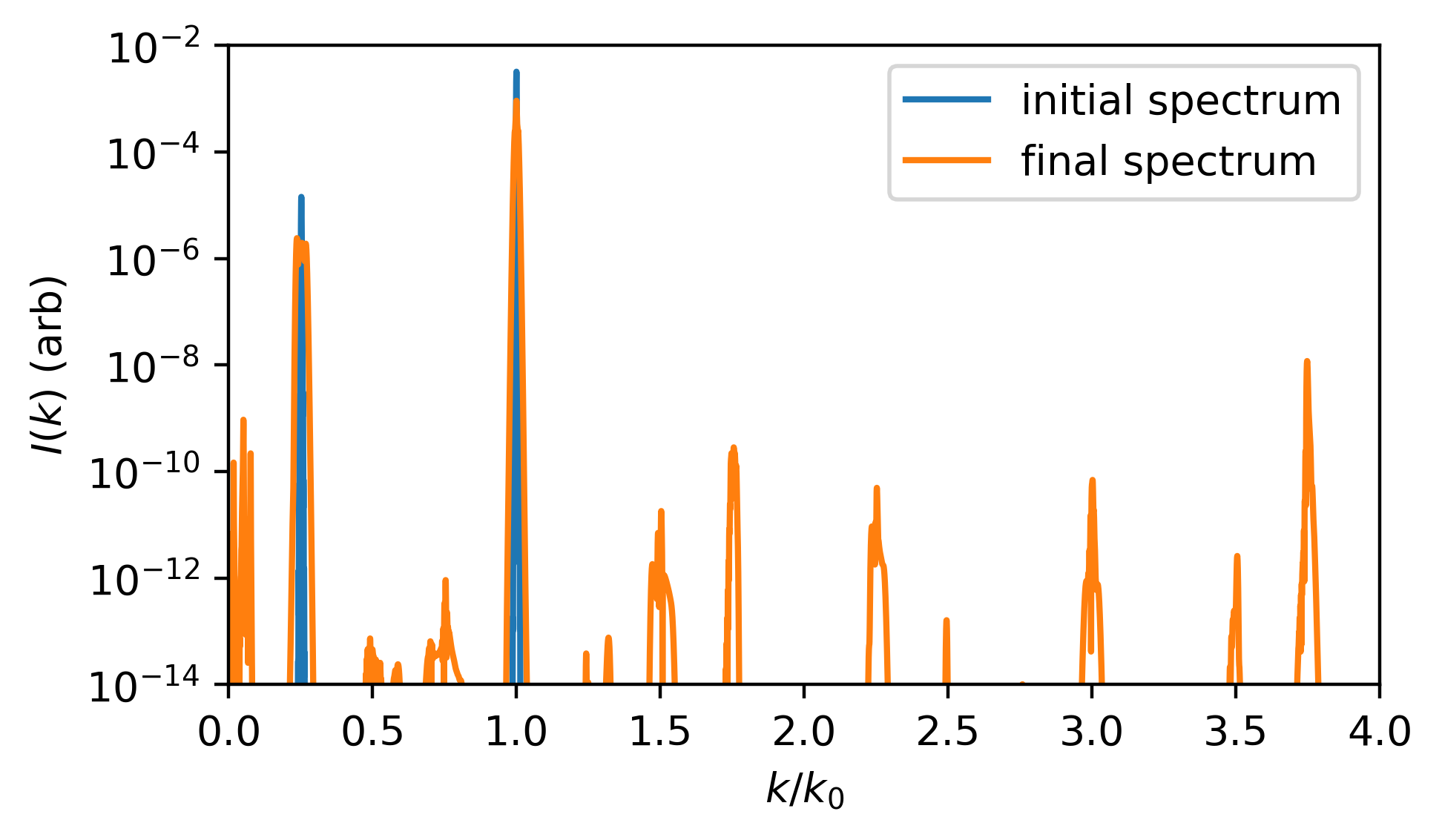}
    \includegraphics[width=\columnwidth]{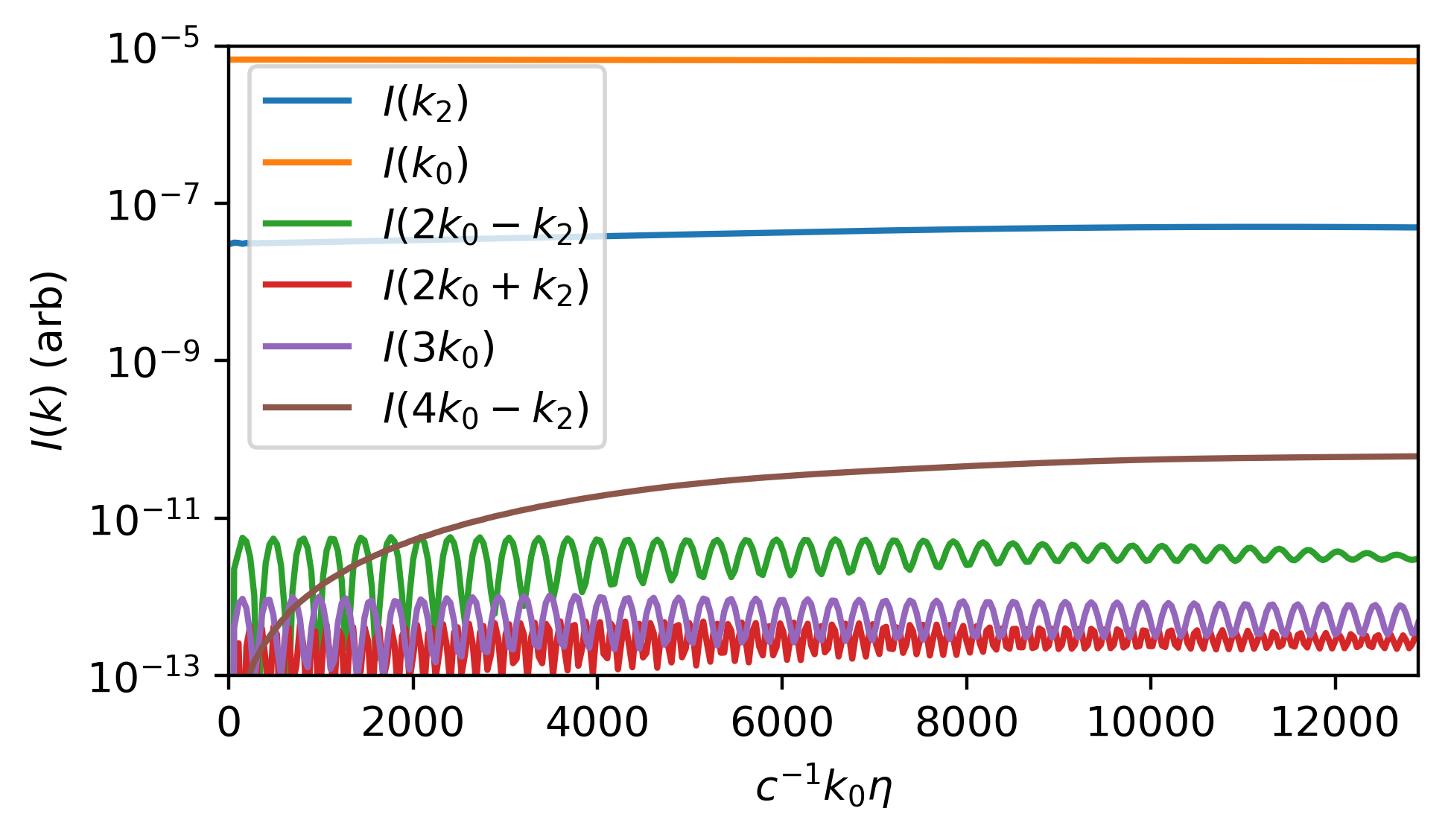}
	\caption{The evolution of the spectrum of a pump ($k_0=1$, $a_0 = 0.5$) and seed ($k_2\approx(2-\sqrt{3})k_0$, $I_2 = I_0/200 $). Both waves are initialized within a homogenous plasma and evolved until a majority of the seed has slipped past the pump. Nonlinear distortions of the pump and seed are weak, but many peaks can be seen in the final spectrum compared to the initial spectrum (top). In $k$ space many spontaneous, but non-growing, modes are instantaneously excited across the spectrum. The particular mode of interest, at $4k_0-k_2$ may grow if Eqn. \eqref{eqn:resonance} is satisfied allowing it to achieve higher intensities than those at modes resulting from lower order processes. This can be seen through looking at the energy in each frequency band, (bottom), where only the resonant mode is growing, and other nonlinearly driven bands oscillate in energy. The energy for each band is calculated by integrating the spectrum in a window of each pick of width $k_{pe}/4$.}
	\label{fig:case_1}
\end{figure}

The nature of this resonance can be demonstrated by varying the wavenumber, $k_2$, of a low frequency seed, and examining the energy which is scattered into the expected resonance at $4k_0 -k_2$.
Both waves start in a homogenous plasma, and are evolved until a minority of the seed overlaps with the pump, as diagramed in the co-propagating coordinate scheme in Fig. \ref{fig:diagram}.
The plasma is initialized such that $k_0/k_{pe} = 7.5$.
While the plasma density is such that it is relatively underdense for the pump wave, we note that the large frequency separation between the pump and the seed wave means the plasma density is not much smaller than the quarter critical density of the seed wave.
The pump is given an $a_0 = 0.5$, and the seed is overlapped with a factor of $200$ lower intensity.
The pulse length, which is set by the slippage condition between the pump and the seed is 450 wavelengths, which propagates for a duration of 4000 laser periods.
The seed frequency is varied around the proposed resonance, with a minimum frequency bounded by two plasmon decay $\omega_2 = 2\omega_{pe}$, and the maximum frequency bounded by the coupling of the third harmonic of the low frequency seed to the stokes shifted pump wave, where $3\omega_2 = \omega_0-\omega_{pe}$.
Both processes at the bounds are lower order and quickly produce a large electron plasma wave, whose dynamics dominate the desired higher order upshift.

The evolution of the simulation spectrum shows the many we nonlinear processes at play, but the resonant six-photon process is the most significant.
The spectrum created by the wave simulation for a resonant simulation is shown in Fig. \ref{fig:case_1}.
For a resonant $k_2$, the strongest growth is of the six-photon resonance, however, energy may be transferred to many other modes.
For example, the second most defined new peak in the spectrum is the third harmonic of the pump wave.
However, even though this mode is created by a lower order and thus more strongly coupled, it is not resonant so it cannot continually grow and is saturated at this level.
The evolution of the energy in different linear combinations of the pump, seed, and plasma frequency shown in Fig. \ref{fig:case_1} demonstrates this.
Other bands oscillate in energy, while the high frequency six-photon resonance grows to a level higher than that of other nonlinear processes.
Many other peaks corresponding to combinations of the pump and seed wave occur in the spectrum, but at weaker levels.
Of particular note is the four-photon process, corresponding to $k=2k_0 - k_2$, experiences some scattering, but is not resonant in the colinear geometry without further tuning of the pump and seed wave.~\cite{malkin_resonant_2020-1}

\begin{figure}[t]
    \includegraphics[width=\columnwidth]{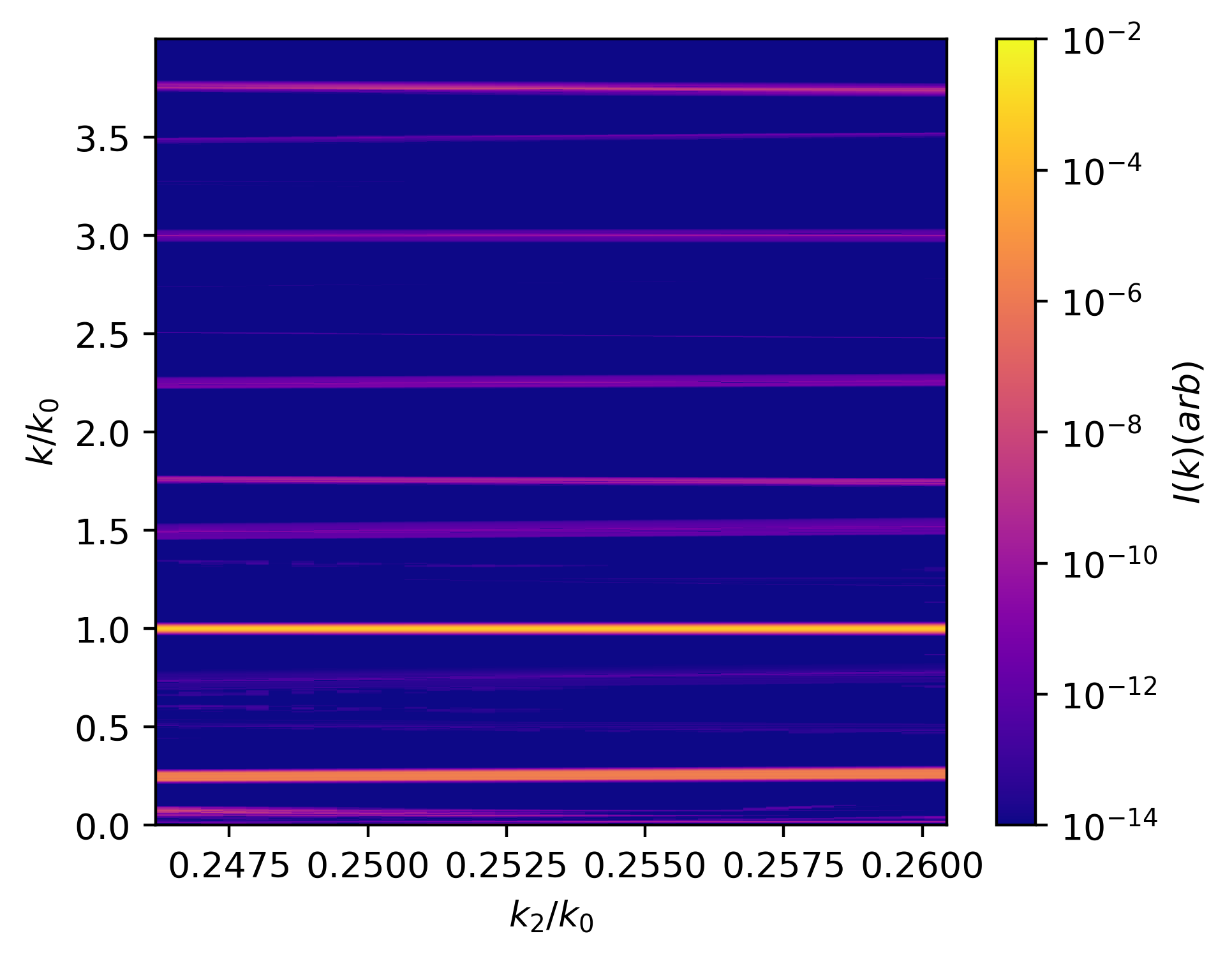}
    \includegraphics[width=\columnwidth]{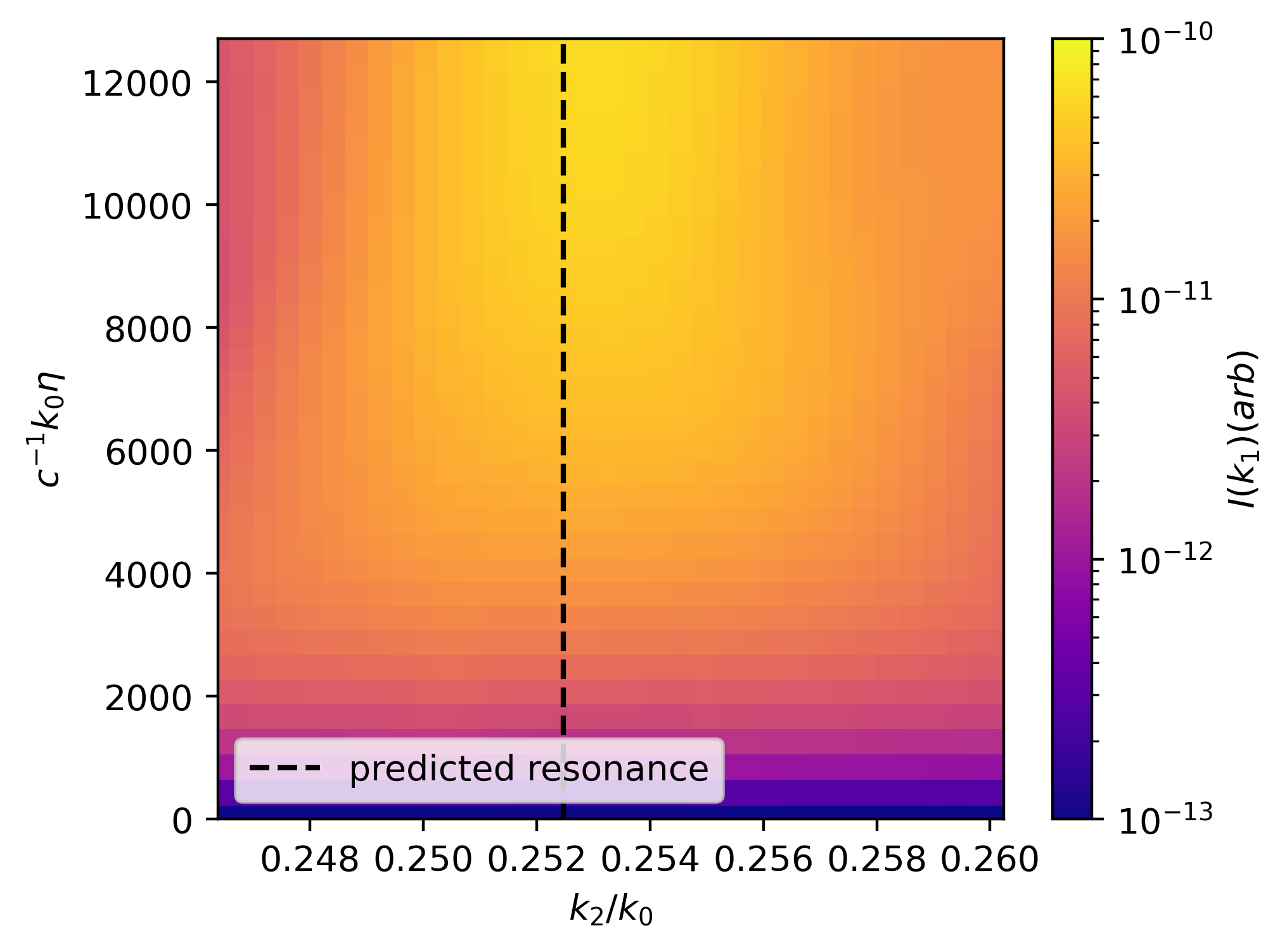}
	\caption{A parameter scan across seed wavenumber $k_2$ results in the varied excitation of many modes, as can be seen in the comparison of the final spectrum by seed frequency shown at top.
	Each vertical slices corresponds to the final spectrum (e.g. Fig. \ref{fig:case_1}) of a corresponding simulation at the specified $k_2$.
	The highest band at wavenumber $k_1=4k_0-k_2$ corresponds to the desired process, and the temporal evolution at varied $k_2$ is shown at bottom.
	The expected resonance at $k_2 = 4k_0-\left[\left(2\sqrt{1+n} -\sqrt{3}\right)^2 - n\right]^{1/2}k_0 = 0.252k_0$ is more highly amplified than neighboring wavenumbers.
	The range of the parameter scan over $k_2$ is heavily restricted by the growth of electron plasma waves at the bounds of $k_2$.
	Shorter duration simulations highlight a larger difference in the energy of the amplified across a wider range of $k_2$, at the cost of less amplification of the $k_1$ wave.}
	\label{fig:scan}
\end{figure}

Varying the seed frequency demonstrates the uniqueness of the resonance point.
Fig. \ref{fig:scan} shows the post interaction spectrum for simulations at varying seed frequency, $k_2$.
The spectra across differing $k_2$ contains many bands.
Flat integer bands correspond to harmonics of the pump.
Those of non-zero slope correspond to modes involving the seed frequency.
The pump and seed wave are the strongest, followed by the top band which corresponds to the $4k_0 - k_2$ mode.
As the seed frequency approaches the right boundary, the third harmonic of the seed approaches the Stokes line of the pump, resulting in significant Raman scattering.
The variation in Fig. \ref{fig:scan} of the high frequency $4k_0 - k_2$ mode is suppressed due to the many orders of magnitude required in the color scale to observe the full range of created modes.

The unique resonant condition at $k_2 = 2\omega_0 - \sqrt{3}k_0$ is more straightforwardly seen in Fig. \ref{fig:scan}.
Some initial small amount of energy can be transferred into the $4k_0 - k_2$ mode regardless of the seeded $k_2$.
However, this energy will oscillate if $k_2$ is not chosen correctly, which occurs as the seed frequency is moved away from the resonance, shown by the dashed line.
As $k_2$ becomes closer to the resonance, the period of oscillation grows, and the energy may coherently grow.
The efficiency of this process remains low, with the energy in the high frequency band more than $10^{-5}$ below the initial pump energy.

Other resonances may result in growing parasitic modes and become limiting.
The desired scattering becomes dominated by Raman scattering if the seed frequency can create an electron plasma wave.
Raman scattering may be seeded off of noise, however, there are two other limits also present as the seed frequency changes.
If the seed frequency approaches the threshold for two plasmon decay at low $k_2$, or the third harmonic of the seed approaches the Stokes shifted frequency of the pump, Raman scattering may becomes strong.
At these modes the spectra starts to produce many $\omega_{pe}$ shifted bands, which distort the spectrum.
When these two resonances occur, the electron plasma wave, and corresponding density fluctuations, quickly grow beyond the limits of the cold hydrodynamic model.
We aim to avoid Raman scattering, so this limitation should not be relevant for $k_2$ detuned from these operating points.
However, at the high density these bounds may be close to the resonance where $\omega = 4\omega_0 - \sqrt{3}k_0$.
This limits the plasma density, and thus the coupling of the six-photon coupling.
Fluctuations in plasma density may push the low frequency wave into resonance and produce a large electron plasma wave.
A noisier environment than the shown simulations will also provide a more fecund environment for parasitic Raman scattering.
Further manipulation of the plasma and laser pulses will be needed such that Raman scattering does not dominate any desired higher order nonlinearity.

\section{Summary and Discussion}
\label{sec:conclusion}
This paper demonstrates through numerical simulation the unique non-integer harmonic resonance driven by six-photon coupling in a plasma.
This coupling, when at resonance, driven by the fifth order corrections from the electron Lorentz factor, is large enough that in the post interaction spectrum the mode is larger than other nonlinear processes.
This is true even though Raman scattering and relativistic four-wave mixing are lower order processes.
It is isolated in low noise pseudospectral simulations.
However, for the pulse durations and propagation distances considered it is still below the strength of the injected waves.
The non-integer multiple of the frequency is required for the six-photon scattered wave to be strong.
If the frequency is detuned, growth of the high frequency mode is orders of magnitude lower, and if it is significantly detuned the injected wave can drive other parasitic processes.

The six-photon process is distinguishable at demonstrated parameters, but reaching efficient upconversion will require further modeling.
The pulses in this paper have a picosecond duration and propagate through 2mm of plasma, producing a high frequency wave of $10^{-5}$ lower intensity than the pump.
This is highly inefficient, but, given the significant pump energy, might be detectable in a sensitive experiment.
However, it falls short of the requirements for application, where scaling the growth length results in a plasma channel length that is unachievable.
This is primarily because a lower plasma density has been used than in the most promising upconversion schemes.~\cite{malkin_six-photon_2023}
The use of high densities and high growth has been limited by the development of plasma instabilities which might be overcome in further simulation.
Inhomogeneous plasma conditions which disrupt the growth of plasma waves, could allow density, and thus coupling to be much higher.
Even in a higher coupling regime, pulses will propagate for many more laser cycles.
Resolving the shortest wavelengths and highest frequencies would be costly, but can be overcome by working with an envelope model, which would only need to resolve envelope length scales and nonlinear process timescales.

Working around these challenges to model, and eventually experimentally produce a high energy, short wavelength pulse could be done through the application of a more sophisticated approach.
Notably, the six-photon resonance here which uses a single pump has a much lower growth rate than those consisting of detuned pump waves.
When the beating of the pumps is chosen precisely, the resonance achieves a smaller frequency multiplication, but with much higher growth.~\cite{malkin_six-photon_2023}
Parasitic Raman scattering might also be suppressed through varying the density of the plasma.
Varying the density would be favorable, as the six-photon resonance considered here is less sensitive to the plasma density than the Raman resonance.
Though it is important to note that this will not remain the case if the beat wave between pumps becomes relevant for other six-photon processes.
Higher order process require a high intensity over a large propagation distance, so filamentation may become a problem if low diffraction and thus high power is needed.
If these conditions are met, simulations performed using particle-in-cell methods could directly show the six-photon process taking into account kinetic effects.
Six-photon scattering contains great promise, and iterated scattering, which might be automatically seeded, could produce geometric increases in frequency.

\begin{acknowledgments}
  The authors thank Vladimir Malkin for informative discussions. 
  This research was supported by NSF PHY-2206691 and DOE DE-SC0021248.
\end{acknowledgments}

\bibliography{six_wave_paper}

\end{document}